\documentclass[usenatbib]{mn2e}

\usepackage{graphicx}
\usepackage{epsfig}
\usepackage{color}
\usepackage{amssymb}
\usepackage{lscape}
\usepackage{float}
\usepackage{amsmath}
\usepackage{color}
\usepackage{longtable}
\usepackage{journal_names}
\usepackage{rotating}
\usepackage{pdflscape}
\usepackage{dcolumn}
\usepackage{caption}
\usepackage{subcaption}
\usepackage{bold-extra}
\usepackage{url}

\usepackage[nameinlink]{cleveref}

\def\la{\mathrel{\hbox{\rlap{\hbox{\lower4pt\hbox{$\sim$}}}\hbox{$<$}}}}
\def\ga{\mathrel{\hbox{\rlap{\hbox{\lower4pt\hbox{$\sim$}}}\hbox{$>$}}}}

\def\arcmin{\hbox{$^\prime$}}
\def\arcsec{\hbox{$^{\prime\prime}$}}

\newcommand{\dg}{^{\circ}}

\newcommand{\MSUN}{${\rm M}_\odot$}

\newcommand{\kms}{{\,km\,s$^{-1}$}}

\newcommand{\HI}{\mbox{\normalsize H\thinspace\footnotesize I}}

\def\apjl   {{ApJL }}
\def\apjs   {{ApJS }}

\title[GASP XVII. HI imaging of the jellyfish galaxy JO206.]{GASP XVII. HI imaging of the jellyfish galaxy JO206: gas stripping and enhanced star formation.}

\author[M. Ramatsoku et al.]{M. Ramatsoku$^{1}$\thanks{E-mail: ramatsoku.mpati@inaf.it}, P. Serra$^{1}$, B. M Poggianti$^{2}$, A. Moretti$^{2}$, M. Gullieuszik$^{2}$,
\newauthor
D. Bettoni$^{2}$, T. Deb$^3$, J. Fritz$^4$, J. H. van Gorkom$^5$, Y. L., Jaff\'{e}$^6$, S. Tonnesen$^7$, 
\newauthor
M. A. W Verheijen$^{3,11}$, B .Vulcani$^{2}$, B. Hugo$^{8,10}$, G. I. G. J\'{o}zsa$^{8,9,10}$, 
\newauthor
F.M. Maccagni$^{1}$, S. Makhathini$^{10,8}$, A. Ramaila$^{8,10}$, O. Smirnov$^{10,8}$, K. Thorat$^{8,10}$.\ \\\\
$^{1}$INAF- Osservatorio Astronomico di Cagliari, Via della Scienza 5, I-09047 Selargius (CA), Italy\\
$^{2}$INAF- Osservatorio Astronomico di Padova, Vicolo dell'Osservatorio 5, I-35122 Padova, Italy \\ 
$^3$Kapteyn Astronomical Institute, University of Groningen, Landleven 12, 9747 AV Groningen, The Netherlands\\
$^4$Instituto de Radioastronomia y Astrofisica, UNAM, Campus Morelia, A.P. 3-72, C.P. 58089, Mexico\\
$^5$Department of Astronomy, Columbia University, Mail Code 5246, 550 W 120th Street, New York, NY 10027, USA\\
$^6$Instituto de F\'{i}sica y Astronom\'{i}a, Facultad de Ciencias, Universidad de Valpara\'{i}so, Avda. Gran Bretana 1111 Valpara\'{i}so, Chile\\
$^7$Center for Computational Astrophysics, Flatiron Institute, 162 5th Ave, New York, NY 10010, USA\\
$^8$South African Radio Astronomy Observatory, 2 Fir Street, Black River Park, Observatory, 7925, South Africa\\
$^9$Argelander-Institut f\"{u}r Astronomie, Auf dem H\"{u}gel 71, D-53121 Bonn, Germany\\
$^{10}$Department of Physics \& Electronics, Rhodes University, Makhanda, 6139, South Africa\\
$^{11}$National Centre for Radio Astrophysics, Tata Institute of Fundamental Research,  Postbag 3, Ganeshkhind, Pune 411 007, India\\\
}

\begin{document}

\date{Accepted 2019 June 3. Received 2019 May 29; in original form 2018 December 21}


\maketitle

\begin{abstract} 
We present VLA \HI\ observations of JO206, a prototypical ram-pressure stripped galaxy in the GASP sample. This massive galaxy (M$_{\ast} =$ 8.5 $\times$ 10$^{10}$ \MSUN) is located at a redshift of $z =$ 0.0513, near the centre of the low-mass galaxy cluster, IIZw108 ($\sigma \sim575$\,\kms). JO206 is characterised by a long tail ($\geq$90\,kpc) of ionised gas stripped away by ram-pressure. We find a similarly long \HI\ tail in the same direction as the ionised gas tail and measure a total \HI\ mass of $3.2 \times 10^{9}$ \MSUN. This is about half the expected \HI\ mass given the stellar mass and surface density of JO206. A total of $1.8 \times 10^{9}$ \MSUN\ (60\%) of the detected \HI\ is in the gas stripped tail. An analysis of the star formation rate shows that the galaxy is forming more stars compared to galaxies with the same stellar and \HI\ mass. On average we find a \HI\ gas depletion time of $\sim$0.5 Gyr which is about four times shorter than that of ``normal'' spiral galaxies. We performed a spatially resolved analysis of the relation between star formation rate density and gas density in the disc and tail of the galaxy at the resolution of our \HI\ data. The star formation efficiency of the disc is about 10 times higher than that of the tail at fixed \HI\ surface densities. Both the inner and outer parts of JO206 show an enhanced star formation compared to regions of similar \HI\ surface density in field galaxies. The enhanced star formation is due to ram-pressure stripping during the galaxy's first infall into the cluster. 
\end{abstract}


\section{Introduction}\label{intro}
One of the fundamental processes that affects the evolution of galaxies is their star formation activity. Galaxy surveys have shown that the star formation activity decreases with redshift such that galaxies with higher star formation rates (SFR) are more abundant at higher redshifts (\citealp{Noeske2007}, \citealp{Madau2014}, \citealp{vanderWel2014}). The gradually decreasing SFR has resulted in passive or  ``quenched'' galaxies and the evolution of galaxies from late- to early-type (\citealp{Dressler1997}, \citealp{Postman2005}, \citealp{Smith2005}). Since gas is the main ingredient in star formation it is crucial to understand how galaxies acquire and lose their gas (\citealp{Larson1972}, \citealp{Dekel2006}, \citealp{Silk2012}). Internal and external processes are known to be involved in the gas acquisition or loss. Of the internal processes, the basic is the cooling of the hot gas in the dark matter haloes of galaxies. As the gas cools it drops down into the interstellar medium (ISM) of the galactic discs (\citealp{White1991}, \citealp{King2015}). This gas sometimes gets mechanically displaced or ionised thus affecting the star formation activity. 

The environment plays a crucial role in creating the external processes which affect the galaxy star formation activity. For instance, galaxies in densely populated environments such as galaxy clusters have been observed to have their star formation dampened or quenched more efficiently than their counterparts in the field (\citealp{Dressler1980}, \citealp{Cooper2007}, \citealp{George2011}, \citealp{Whitaker2012}, \citealp{Nantais2017}, \citealp{Foltz2018}). This is often due to gas removal mechanisms that take place in these dense environments (\citealp{Boselli2006}, \citealp{DeLucia2011}). 

In processes such as mergers gas is removed from a galaxy as it becomes gravitationally detached during the collision (\citealp{Toomre1972}, \citealp{Walker1996}). Depending on the geometry and total angular momentum of the system it is possible for the perturbed gas to be funnelled to the nuclear regions where it sometimes gets consumed by star-formation or contributes to fuelling an Active Galactic Nuclei (\citealp{Baldry2004}, \citealp{Balogh2009}). In processes such as harassment the gas and stellar distributions get perturbed. In instances of this process the majority of the gas clouds will be affected and the galaxy will undergo an abrupt burst of star formation which consumes all of the fuel for new stars (\citealp{Moore1996}, \citealp{Duc2008}, \citealp{Smith2015}). In less severe harassment cases only the diffuse gaseous halo is perturbed, effectively stopping the gas from cooling and condensing thus quenching star formation in a galaxy (\citealp{Dressler2013}, \citealp{Cattaneo2015}, \citealp{Peng2015}, \citealp{Jaffe2016}). 

Other mechanisms only affect the gas component of galaxies. One of the most well-known is ram-pressure stripping. This particular process has often been observed in rich galaxy clusters containing hot X-ray emitting gas which forms the intra-cluster medium (ICM). As a galaxy falls into the cluster core it passes through this ICM which exerts hydrodynamical pressure on the galaxy. If the ICM pressure is sufficiently high it can overpower the gravitational force keeping the gas bound to the galaxy, effectively stripping the galaxy of its star forming fuel (\citealp{Gunn1972}, \citealp{Moran2007}, \citealp{Porter2008}, \citealp{Dressler2013}, \citealp{Jaffe2015}).  

Extreme examples of ram-pressure stripping are seen in the so called ``jellyfish'' galaxies. These objects
are so named because they show one-sided tails seemingly stripped from the main galaxy body (\citealp{Yagi2007}, \citealp{RSmith2010}, \citealp{Fumagalli2014}). Knots of recent star formation are often observed in these tails.

The GAs Stripping Phenomena survey (GASP;  \citealp{Poggianti2017}) was conducted with the aim of identifying and collecting a statistically significant sample of these galaxies in nearby clusters ($z = 0.04 - 0.07$) from the WIde-field Nearby Galaxy-cluster Survey (WINGS; \citealp{Fasano2006}). Within the context of the GASP sample, a galaxy is considered a jellyfish if its H$\alpha$ tail is at least as long as its stellar disc diameter. Using the MUSE Integral Field spectrograph on the VLT, 94 optically selected stripping candidates \citep{Poggianti2016} were observed. This sample of galaxies has a wide range of jellyfish morphological asymmetries and masses in varying environments. The focus of the survey is to study all phases of the gas and stellar populations in these galaxies, and to quantify the amount of star formation activity during the process of gas stripping.

One of the primary ways to examine the star formation activity of a galaxy in relation to gas stripping is by studying its neutral gas (\HI) content. A total of about 25 per cent of the interstellar medium of a typical spiral galaxy is composed of \HI\ \citep{Boulares1990}. It is usually distributed out to large radii of about 1.5 - 1.8 optical $R_{25}$ radii (\citealp{Broeils1997}, \citealp{Walter2008}), where the gravitational force binding it to the host galaxy is weaker, thus making it easy to remove, particularly in cluster and group environments (\citealp{Haynes1984}, \citealp{Bravo-Alfaro1997}, \citealp{Gavazzi2008}). This \HI\ property makes it an excellent tracer of tidal or hydrodynamical gas removal processes (\citealp{Kapferer2009}, \citealp{Chung2009}, \citealp{Abramson2011}, \citealp{Jaffe2015}, \citealp{Yoon2017}).

\HI\ observations of spiral galaxies in clusters such as Virgo \citep{Chung2009} have shown that galaxies located at low cluster-centric distances ($\leq$0.5\,Mpc) tend to have smaller \HI\ discs than the stellar disc \citep{Chung2007}. These results are often explained by ram-pressure stripping. In the Virgo cluster, the VLA Imaging of Virgo in Atomic gas survey (VIVA; \citealp{Chung2009}) found that 7 spirals galaxies out of a sample of 50 had long \HI\ tails extending well beyond the optical disc. All of these tails point away from the centre of the cluster (M\,87).  \HI\ tails were also reported in the Coma cluster through \HI-imaging observations of 19 brightest galaxies in the cluster \citep{BravoAlfaro2000}. In addition to asymmetries in \HI, tracers of young stars in the form of H$\alpha$ and UV emission were also found to be a typical feature of ram-pressure stripped galaxies (\citealp{Cortese2006}, \citealp{Sun2007}, \citealp{Kenney2008}, \citealp{Yagi2010}, \citealp{Fumagalli2014}, \citealp{Boselli2016}, \citealp{Fossati2016}). In rare cases tails are observed in radio continuum and X-rays as well (\citealp{Gavazzi1995}, \citealp{Sun2005}, \citealp{Vollmer2004}, \citealp{Sun2010}).

There is a large body of simulation-based work studying ram pressure stripping (e.g. \citealp{Abadi1999}, \citealp{Quilis2000}, \citealp{Schulz2001}, \citealp{Roediger2005}, \citealp{Roediger2006}, \citeyear{Roediger2007}, \citeyear{Roediger2008}; \citealp{Jachym2007}, \citeyear{Jachym2009}; \citealp{Tonnesen2009}, \citeyear{Tonnesen2010}; \citealp{Vollmer2001}, \citeyear{Vollmer2003}, \citeyear{Vollmer2006}, \citeyear{Vollmer2008}, \citeyear{Vollmer2009}, \citeyear{Vollmer2012}; \citealp{Kapferer2009}, \citealp{Kronberger2008a}) some of which has focused on \HI.  For example, Vollmer et al. (\citeyear{Vollmer2003},  \citeyear{Vollmer2005}, \citeyear{Vollmer2006}, \citeyear{Vollmer2008}) uses N-body simulations and models of orbiting galaxies in the Virgo cluster to carefully model both the remaining and stripped \HI\ gas.  Through comparisons with observations, the authors interpret the stripping and interaction history of several galaxies in the cluster (see also \citealp{Merluzzi2013},\citeyear{Merluzzi2016}, \citealp{Gullieuszik2017} for similar comparisons with H$\alpha$-emitting gas). The \HI\ velocity and spatial information of of ram pressure stripped galaxies may allow for tight constraints on the history of satellite galaxies.

Simulations have also been used to make more general predictions about ram pressure stripping.  For example, \citet{Quilis2000} argued that observed holes in the \HI\ gas disk allowed for fast, complete ram pressure stripping. Using smooth gas density profiles, \citet{Roediger2007} find that the amount of gas stripped is well matched by the analytical arguments of \citet{Gunn1972}, that gas will be removed at radii where ram pressure overcomes the restoring force of the disk. Using simulations that included radiative cooling, \citet{Tonnesen2009} also found that the total gas removed from galaxies was similar to that predicted using the \citet{Gunn1972} prescription, but stripping acted much more quickly. The ICM wind removed low-density gas from a range of radii, leaving behind dense clouds (also see \citealp{Schulz2001}).  \citet{Tonnesen2010} focused on the stripped tail of gas, and found that radiative cooling allowed for long, narrow tails of stripped gas with high surface-density \HI, as observed in some ram pressure stripped galaxies.

To fully understand the effect of ram-pressure stripping on the GASP galaxy sample it is imperative to examine the \HI\ content and distribution of these galaxies. 

In this paper we focus on the \HI\ gas phase of a quintessential ``jellyfish'' galaxy in the GASP sample, namely JO206 ($\alpha_{J2000}, \delta_{J2000}$, $z$  = 1:13:47.4, +02:28:35.5, = 0.0513; \citealp{Gullieuszik2015}, \citealp{Moretti2017}). This massive galaxy is located near the centre of a low-mass galaxy cluster and exhibits a long tail ($\geq$90\,kpc) of ionised gas stripped away by ram-pressure \citep{Poggianti2017}. Our aim is to investigate the impact of this stripping event on the neutral ISM of JO206, and to understand the relationship between this neutral ISM and the star formation activity traced by the ionised gas both within and outside the galaxy disc.

In section\,\ref{environs} we give an overview of the JO206 properties and the environment in which it resides. A further brief discussion of the galaxy's currently available multiwavelength data is given in the same section. \HI\ observations conducted with the VLA and data processing are outlined in section\,\ref{sec:vlaObs}. We provide an analysis of the \HI\ results in section\,\ref{HIresults}. In section\,\ref{coldSFR} we assess and discuss the relation between the \HI\ gas and star formation activity in the galaxy. The analyses and discussions are summarised in section\,\ref{summary}.

Throughout this paper, we adopt a \citet{Chabrier2003} Initial Mass Function (IMF) and assume a $\Lambda$ cold dark matter cosmology with $\Omega_{\rm M} = 0.3, \Lambda_{\Omega} = 0.7$ and a Hubble constant, H$_{0}$ = 70 \kms\ Mpc$^{-1}$.

\section{JO206 properties and Environment}\label{environs} 
The JO206 galaxy is massive with a total stellar mass of $8.5 \times10^{10}$ \MSUN\ and hosts an active galactic nucleus (AGN) (\citealp{Poggianti2017nat}). It is a member of a poor galaxy cluster known as IIZw108 ($z$ = 0.04889; \citealp{Biviano2017}) in the WINGS/OmegaWINGS sample (\citealp{Fasano2006}, \citealp{Cava2009}, \citealp{Varela2009}, \citealp{Gullieuszik2015}, \citealp{Moretti2017}). The host cluster has a velocity dispersion and an Xray luminosity of $\sigma_{cl} \sim\ 575 \pm 33$\kms\ (\citealp{Biviano2017}) and $L_{\rm X} = 1.09 \times 10^{44}$ erg.s$^{-1}$ (0.1 - 2.4 keV; \citealp{Smith2004}), respectively and a dynamical mass M$_{200} \sim 2 \times 10^{14}$\MSUN\ \citep{Biviano2017} 

The galaxy has been assigned the highest jellyfish morphological classification of JClass = 5 since it exhibits the most recognisable tail of debris material that is apparently stripped from the galaxy main body \citep{Poggianti2016}. The stripped tail of material is thought to be the result of ram-pressure stripping due to the ICM of the IIZw108 galaxy cluster. This claim has been supported by the galaxy location close to the cluster centre at the projected radial distance of $\sim$350\,kpc (see Fig.\,\ref{vband}) and its high line-of-sight velocity of $1.5\sigma_{cl}$ relative to the cluster's mean systemic velocity and centre. 

\subsection{Estimated fraction of stripped gas}\label{estGasFrac}

Analyses of the dynamics of IIZw108 have shown that JO206 does not belong to any substructure and appears to be an isolated galaxy falling into the cluster \citep{Poggianti2017}.
We re-calculated the fraction of stripped gas presented in \citet{Poggianti2017} using both the truncation radius in H$\alpha$ and the location of JO206 in projected position vs. velocity phase-space, following the method presented in \citet{Jaffe2018}, and using the beta-model for IIZw108's ICM presented in \citet{Reiprich2001} (rather than the Virgo cluster). An in-depth description and detailed calculation of the ram-pressure stripping strength of the cluster are discussed and presented in Sect.\,7.6 in \citet{Poggianti2017}; see also \citet{Jaffe2018}.
In short, we assume instantaneous gas stripping \citep[following][]{Gunn1972} from a pure exponential disk falling into an homogeneous and symmetrical ICM. 
In the phase-space method we compare the intensity of ram-pressure at the projected position and velocity of the galaxy  within the cluster ($P_{\rm ram} \simeq 5.3 \times 10^{-13} N m^{2}$) with the restoring force of the galaxy ($\Pi_{\rm gal}$ which decreases with radial distance from the centre of the galaxy). Stripping will occur when $P_{\rm ram} > \Pi_{\rm gal}$. 
Using the galaxy radius where the stripping condition is found, we then estimate the amount of gas lost due to stripping to be $\sim$70\% (assuming an initial gas fraction  of 10\% with an extent 1.7 larger than the stellar disk). 
As this method has a lot of uncertainties \citep[see caveats section in][]{Jaffe2018}, we also estimated the amount of gas lost purely from the extent of H$\alpha$ relative to the size of the disk, and get a lower fraction of stripped gas ($\sim$40\%). 
We therefore conclude that J0206 has lost between $\sim$40\% and $\sim$70\% of its total gas mass due to ram-pressure stripping.

\subsection{Star formation}
Regardless of this gas loss, ram-pressure stripping is believed to have resulted in a burst of new stars. The galaxy has a reported star formation rate of 5.6 \MSUN\ yr$^{-1}$ determined from H$\alpha$ and excluding the contribution of the central AGN \citep{Poggianti2017}. All the star formation rates reported in this paper are computed from the MUSE $\rm H\alpha$ luminosity corrected for dust and stellar absorption using eqn.(1) in Poggianti et al. (2017b), with the dust correction being estimated from the Balmer decrements measured from the MUSE spectra.\footnote{Note that star formation rates are derived adopting a Chabrier (2003) IMF and a \citet{Cardelli1989} extinction curve.} The JO206 stellar spatial distribution shows that the oldest stars with ages $>$ 0.6 Gyr are only found within the main galaxy body. The stripped tail on the other hand comprises young stars that would have started forming $\sim$0.5 Gyrs ago or less (see Fig.\,16 by \citealp{Poggianti2017}).

\begin{figure}
 \includegraphics[width=85mm, height=80mm]{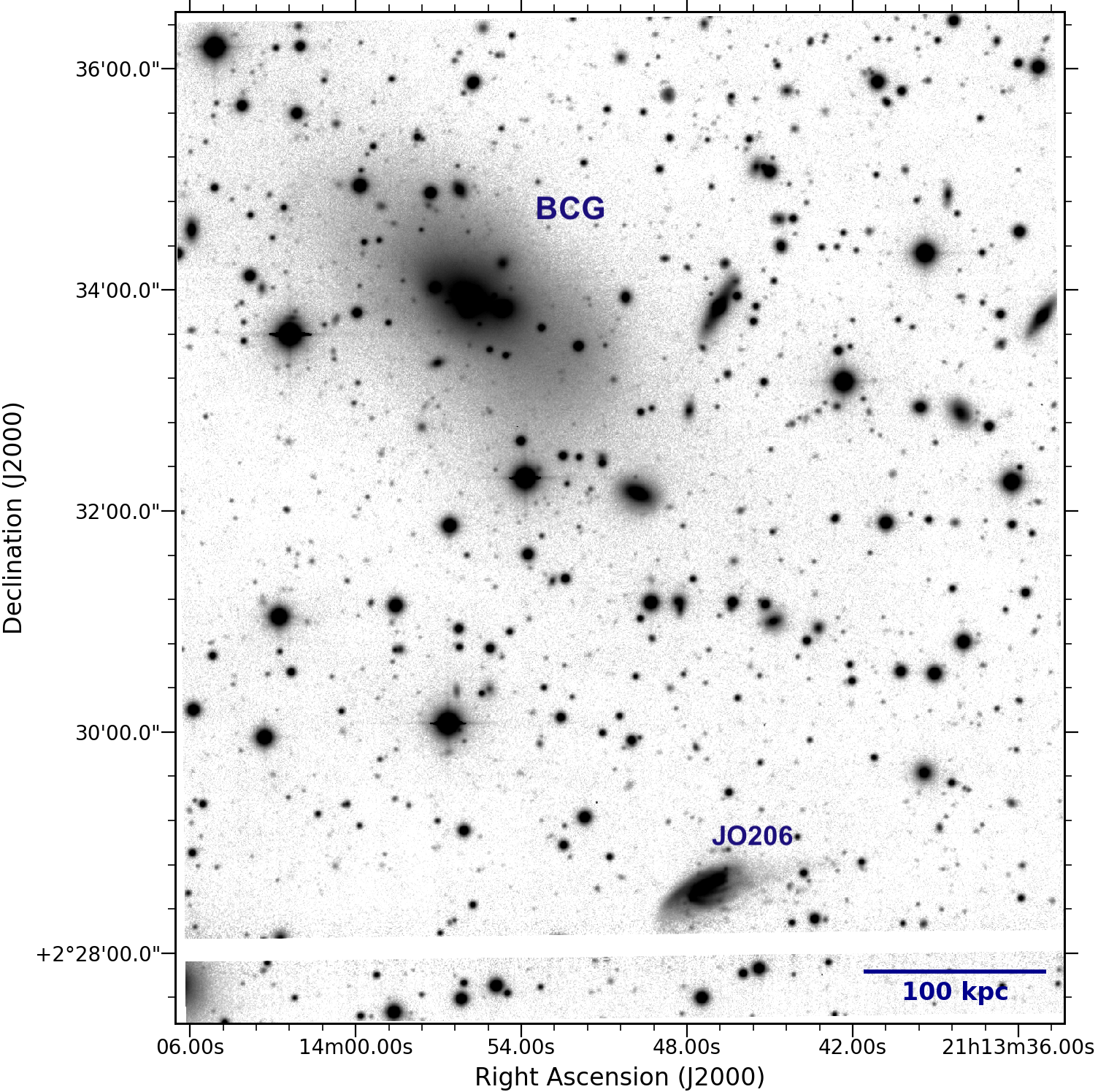}
    \caption{A V-band image of the central region of the IIZw108 galaxy cluster from WINGS. The JO206 galaxy and the BCG that defines the centre of the cluster are labelled as such. They are separated by $\sim$350\,kpc at the distance of the cluster of $\sim$208\,Mpc.}\label{vband}
 \end{figure}

\subsection{MUSE and APEX observations}\label{HIpha} 
The galaxy was first observed with the MUSE IF spectrograph in August 2017. Observations were carried out using two pointings which covered both the galaxy body and tail. The exposure time for each of the pointing was 2700 seconds with a seeing of 1\arcsec\ and 1.2\arcsec. A detailed description of these observations and the data reduction is given in \citet{Poggianti2017}. That paper also presents a detailed analysis of the MUSE observations, that show the $\rm H\alpha$ tentacles of the stripped debris material. We have also obtained recent MUSE observations covering the southern part of the galaxy. These observations were conducted to investigate a possible southern H$\alpha$ tail which was hinted by the \HI\ observations (see Sect.\,\ref{HIresults}). Throughout this paper we use this recent H$\alpha$ map which includes the southern region of the galaxy. The full details of these additional observations and data reduction are given in Appendix\,A.

\begin{figure}
 \includegraphics[width=88mm, height=50mm]{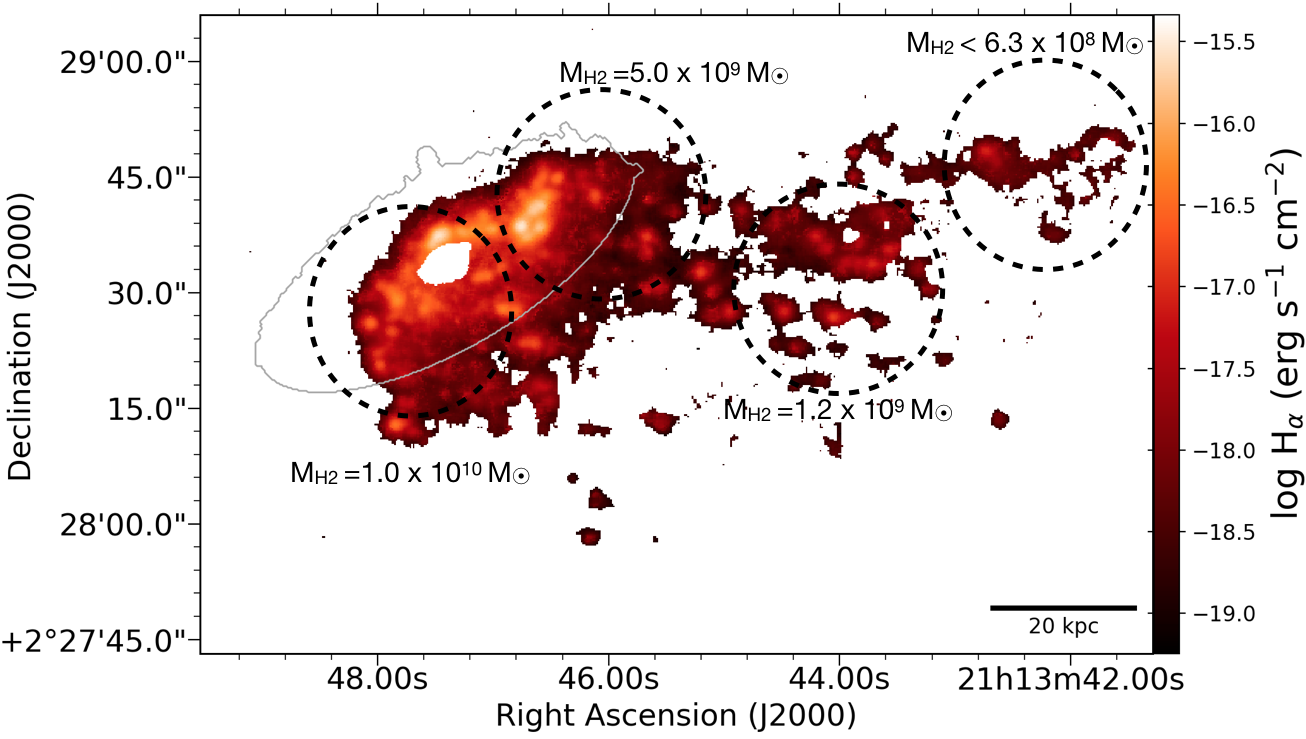}
    \caption{The MUSE H$\alpha$ map \citep{Poggianti2017} with the APEX pointings overlaid \citep{Moretti2018}. The contour delineates the main galaxy body (see the text for a description) and the location of the AGN is represented by the white patch in the middle of the galaxy's main body. The approximate FWHM of the APEX beam are shown the by the black dashed circles with the measured H$_{2}$ mass indicated for each region.}\label{COMap}
 \end{figure}

The tentacle of debris material apparent in the optical image (see Fig.\,\ref{vband}) becomes much clearer in the H$\alpha$ MUSE map shown in Fig.\,\ref{COMap}. In this map the main galaxy body (i.e., the stellar disc) is outlined by the grey contour. The contour was defined from the from the MUSE image using the continuum at the H$\alpha$ wavelength, with the isophote fit at surface brightness of 1$\sigma$ above the average sky background level (Gullieuszik et al., in prep). The image shows a tail of H$\alpha$ emission extending over 90\,kpc to the west of the main galaxy body characterised by regions of clumpy and diffuse H$\alpha$ emission. 

In addition to the optical and H$\alpha$ data, the molecular gas phase of JO206 has also been observed with the Atacama Pathfinder Experiment (APEX; \citealp{Gusten2006}) telescope. These observations were conducted in December 2016 and April - July 2017.  The $^{12}$CO(2-1) transition was observed in four locations of the galaxy shown in Fig.\,\ref{COMap}. The APEX pointings covered the main galaxy body and the tail. The details of these observations, the data reduction and analysis are fully described in \citet{Moretti2018}.

The APEX data revealed clear CO line detections in the central location of the main galaxy body where the AGN is located, and where the tail of the galaxy begins at about 30\,kpc west of the centre. In these two locations the CO emission coincides with the bright H$\alpha$ emissions. However, no secure CO detections coinciding with the H$\alpha$ peaks over 40\,kpc away from the galaxy disc were found \citep{Moretti2018}.

All these data have provided an excellent laboratory comprising the ionised and molecular gas and stars within the galaxy disc and its tails. This is all due to the MUSE large field of view which has allowed for the examination of the galaxy's main body, its tails and surroundings. 

As discussed above, this jellyfish galaxy is experiencing ram-pressure stripping whilst falling into the cluster for the first time. Because of its first infall status it still contains a lot of its gas, thus the recent star formation along its tail. Given the close relation between star formation and cold gas, one of the crucial missing components to complete the picture of the ram-pressure stripping effect on the star formation activity of this galaxy was its neutral gas (\HI) phase.

\section{HI Observations and Data Processing}\label{sec:vlaObs} 
\HI\ observations were conducted with the Very Large Array (VLA; \citealp{Perley2009}) in its C-array configuration between July 2017 and August 2017.  We chose this configuration to achieve a spatial resolution of $\sim$15\,arcsec which would allow the detection of the low surface brightness \HI\ particularly in the galaxy tail. The galaxy field was observed for a total on-source integration time of $2 \times 8$ hours with an additional 4 hours for the bandpass and phase calibrators. Data were obtained over the frequency range of 1335 to 1367\,MHz, centred at 1351 MHz. This covered a total effective bandwidth of 32\,MHz with 1024 channels that are 31.25\,kHz wide (6.56\,\kms\ for \HI\ at $z$ = 0). With this configuration both the JO206 galaxy and the IIZw108 galaxy cluster as a whole were covered by the observed pointing. A summary of the observational parameters is given in Table.~\ref{summaryObs}.

The uv-data were processed following standard procedures using a new data reduction pipeline developed at SARAO\footnote{https://www.ska.ac.za/about/sarao/} and INAF-Cagliari\footnote{http://www.oa-cagliari.inaf.it/} to process mainly data from upcoming MeerKAT\footnote{http://public.ska.ac.za/meerkat/meerkat-large-survey-projects} surveys. It also works on data from other interferometers such as the VLA, GMRT, etc. This pipeline comprises various radio data reduction tools and packages such as CASA \citep{McMullin2007}, AOflagger \citep{Offringa2010}, SoFiA \citep{Serra2015}, WSclean \citep{Offringa2014} among many others. 

Within this pipeline CASA tasks were used to determine and apply antenna-based complex bandpass (independent of time) and gains (independent of frequency). Our calibrators were 3C 48 (for bandpass and flux scale) and J2130+0502 (for the gains). Strong radio frequency interference affecting the calibrators and the target source was flagged using AOflagger. Continuum source subtraction was performed in the uv-plane using the \textit{uvcontsub} task in CASA. This was done in two steps, firstly we made a linear fit to the uv-data using all the channels. The flagged, calibrated and continuum subtracted uv-data were Fourier transformed into an image cube with WSclean. We then used SoFiA to identify channels with \HI-line emission, and repeated the continuum subtraction excluding those channels from the fit. A 3rd order polynomial fit was necessary to subtract all continuum emission satisfactorily. The improved continuum subtracted uv-data were the re-Fourier transformed into a final \HI\ cube using a pixel size of 5 arcsec, a field of view of 0.7 deg$^2$ and natural weighting with \textit{Briggs} robustness parameter $=$ 2 to optimise surface-brightness sensitivity. The resulting cube has an rms noise level of $\sigma \approx$ 0.3 mJy/beam per channel. To eliminate the sidelobes of the synthesised beam we used WSclean and SoFiA iteratively to define 3D clean regions in the \HI\ cube and clean within them down to 0.5$\sigma$. The restoring Gaussian PSF has FWHM of 18 and 26 arcsec along minor and major axis ($\sim$18\,kpc $\times$ 26\,kpc), respectively (PA=159$\dg$). 
Our observational setup allowed us to reach the \HI\ column density sensitivity of $3 \times 10^{19}$ atoms cm$^{-2}$ assuming a line width of 30 \kms\ at the $3\sigma$ noise level. 
To obtain an \HI\ cube at higher angular resolution we repeated the imaging steps with a range of values of the Briggs robust parameter. In Sect.\,\ref{HIresults} we will also show results obtained with robust = 0, which gave the highest obtainable angular resolution of 14\arcsec $\times$ 13\arcsec, a slightly higher noise level of $\sim$0.4 mJy/beam and a column density sensitivity of $9 \times 10^{19}$ atoms cm$^{-2}$ at 3$\sigma$ and for a 30 \kms\ linewidth.

\begin{table}
\caption {A summary of the H\textsc{i} observations.}\label{summaryObs}
\begin{tabular}{llllll}
\hline
 Properties& JO205-206   \\  
  
     \hline\hline
Pointing centre: &     \\
~~~~$\alpha$ (J2000) &$21^{\rm h}13^{\rm m}46^{\rm s}.7$ \\
 ~~~~$\delta$ (J2000)&  $02\dg21\arcmin20\arcsec.0$   \\
Central velocity (radio, barycentric)&15388\,\kms&         \\
Calibrators: & \\
~~~~Gain& J2130+0502                  \\
~~~~Flux and bandpass &   3C48        \\                   
On-source integration & 16 hrs \\  
Observation dates& July \& August 2017\\
Sensitivity (r.m.s per channel) & 0.3 mJy beam$^{-1}$\\
Channel width & 6.56\,\kms \\
Beam (FWHM) (P.A) & $26\arcsec \times 18\arcsec\ $ ($159\dg$) \\ 
\hline\hline\\
\end{tabular}
\end{table}

~\\
\textbf{\HI\ column density and mass:} The \HI-line emission in units of Jy/beam of the galaxy of interest (JO206) was extracted from the image cube using SoFIA. We then converted the \HI\ map units from Jy/beam km/s  to column densities in atoms cm$^{-2}$ using;
\begin{flushleft}
\begin{equation}
\mathrm{N_{\rm HI} = 1.104 \times 10^{21} \int{\frac{S_{\rm v}}{B_{\rm maj} B_{\rm min}}} dv }
\end{equation}
\end{flushleft}

where S$_{\rm v}$ is the flux density in mJy/beam, B$_{\rm maj}$ and B$_{\rm min}$ are the major and minor axes of the beam in arcsec and dv is the channel width in \kms.  

We derived the \HI\ mass (\MSUN) using the formulation, 
\begin{flushleft}
\begin{equation}\label{HImassEq}
\mathrm{M_{\rm HI} = 2.36 \times 10^{5}D^{2}\int S_{\rm v} dv},
\end{equation}
\end{flushleft}
where  $\int$S$_{\rm v}$dv is the total integrated flux expressed in Jy\,\kms\ and D is the distance to the galaxy in Mpc.

\section{Results}\label{HIresults} 
We calculated the total \HI\ mass for JO206 using Eq.\ref{HImassEq} with $\int$S$_{\rm v}$dv $=$ 0.27\,Jy\,\kms\ as derived from the \HI-map. We assume that the galaxy is a the same distance as the cluster, D $=$  208\,Mpc. With these values we measured a total \HI\ mass, M$_{\rm HI} = 3.2 \times 10^{9}$ \MSUN.

\subsection{HI Morphology}

\begin{figure*}
   \centering
 \includegraphics[width=120mm, height=85mm]{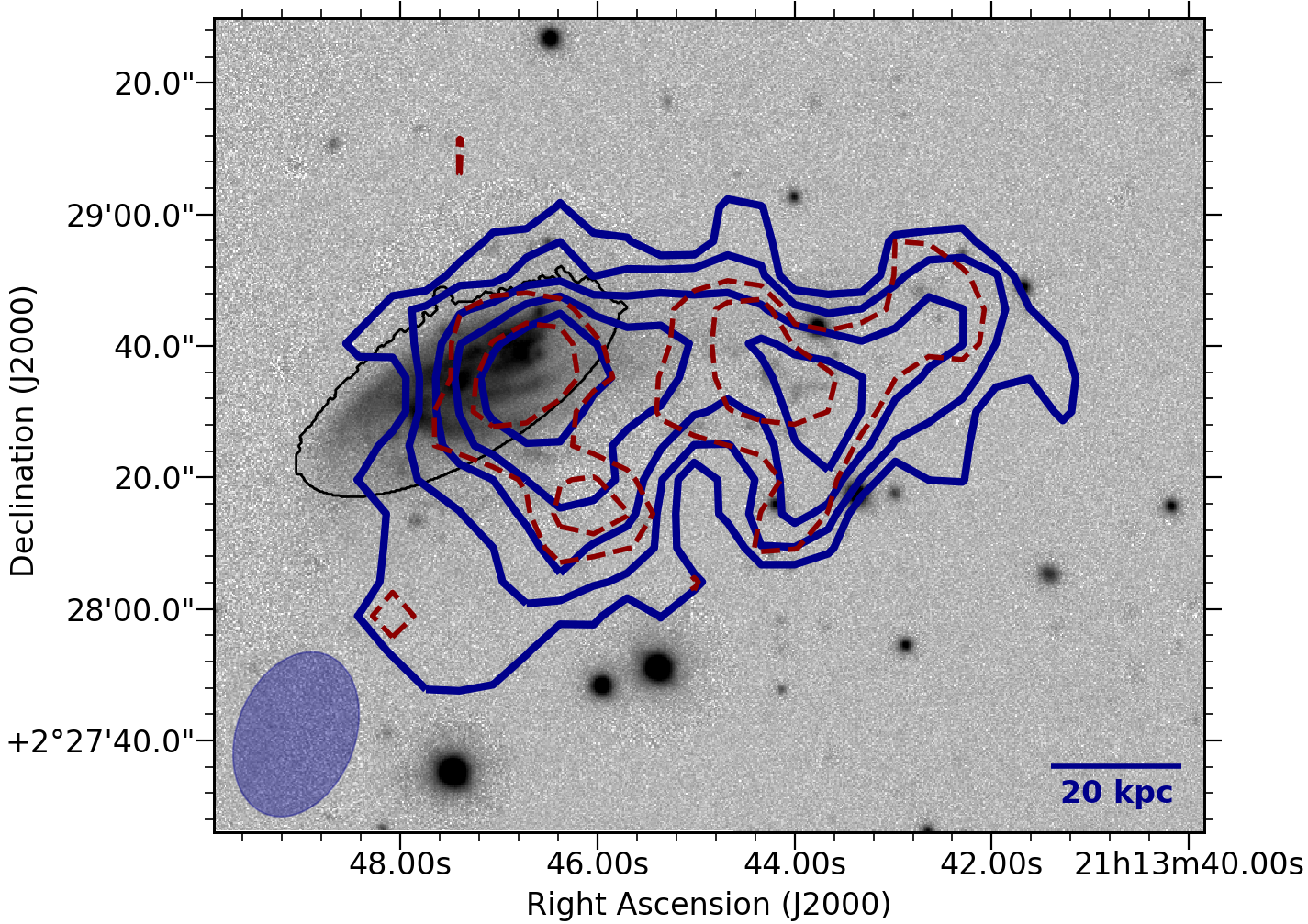}
    \caption{The VLA H\textsc{i} column density contours overlaid on the V-band image of JO206 from WINGS (V-band image from \citealp{Moretti2014}). The blue contour drawn at column densities of 3, 6, 9, ... $\times$ 10$^{19}$ atoms/cm$^{2}$ are from an image with FWHM beam size of $26\arcsec \times 18\arcsec$ indicated by the blue ellipse. The red contours are column densities levels of 10, 20, ... $\times$ 10$^{19}$ atoms/cm$^{2}$ from an image with FWHM beam size of $14\arcsec \times 13\arcsec$. The isophote defining the stellar disc (similarly to Fig.\,\ref{COMap}) is outlined by the black contour.}\label{HIoverVband}
 \end{figure*}

Figure.\,\ref{HIoverVband} shows the \HI\ column density distribution overlaid on an optical V-band image from WINGS. For completeness we also show the contours extracted from an \HI\ cube with an FWHM beam size of 14\arcsec $\times$ 13\arcsec. These are plotted to highlight locations with high density \HI. When compared to the optical image, the \HI\ distribution appears compressed and truncated within the stellar disc on the east side. On the west side it extends well beyond the stellar disc. This extension coincides with the $\sim$90\,kpc H$\alpha$ tail shown in Fig.\,\ref{COMap}. Furthermore, the peak \HI\ surface brightness is 10\arcsec\ ($\sim$10\,kpc) offset from the optical centre of the galaxy. This offset is in the same direction as the \HI\ and H$\alpha$ tails. We have inspected the optical image (Fig.\,\ref{vband}) and the entire \HI\ cube over a spatial area of 0.7 deg$^{2}$ and the radial velocity range of $v_{rad}\sim$12000 - 17000\,\kms, we found no obvious and potentially interacting cluster member(s) near and around JO206 nor along its \HI-tail. Thus, the observed \HI\ morphology is consistent with the effect of ram-pressure gas stripping by the ICM of its host galaxy cluster.

\subsection{HI Deficiency} 
As shown in the preceding section, the \HI\ distribution of JO206 appears to have been affected by ram-pressure resulting in a long \HI\ tail. In this section we examine how much of the galaxy's \HI\ content has been displaced or removed by ram-pressure. This is done by comparing the current, measured \HI\ fraction (M$_{\rm HI}/$M$_{\ast}$) of JO206 with the expected \HI\ fraction based on known \HI\ scaling relations derived from large samples. We use relations by \citet{Brown2015} obtained using ALFALFA data. These relations were chosen because they are based on the \HI\ spectral stacking technique. They are therefore expected to be more robust since they are not biased towards \HI-rich, detected galaxies (\citealp{Huang2012}, \citealp{Fabello2012}, \citealp{Brown2015}). Here we assess the observed \HI\ gas fraction as a function of stellar mass and surface density.

The stellar surface density of JO206 was derived using, 
\begin{equation}
\mathrm{\mu_{\ast} = \frac{M_{\ast}}{2\pi R^{2}_{50}}},
\end{equation}

where M$_{\ast}$ is the stellar mass (see \cref{environs}) and R$_{50}$ is the Petrosian radius containing 50\% of the flux level. We find a stellar mass surface density, $\mu_{\ast} = 4.2 \times 10^{8}$ \MSUN\ kpc$^{-2}$. Note that the scaling relations by \citet{Brown2015} are based on $z$-band R$_{50}$ radii while for JO206 we have used the $K$-band R$_{50}$ radius from \citet{Valentinuzzi2009}; $z$-band parameters are not available for this galaxy. Based on the SDSS (DR15; \citealp{Aguado2019}), 2MASS \citep{Skrutskie2006} and parameters of galaxies in the xGASS sample \citep{Catinella2018}, we find that $z$-band R$_{50}$ radii are systematically smaller than the $K$-band ones by a factor 1.15. This results in a systematic increase of $\mu_{\ast}$ by a factor 1.32, which is small compared to the uncertainty on $\mu_{\ast}$ associated with the M$_{\ast}$ in Eq. 3, and furthermore, does not change the conclusions of our comparison between JO206 and a control sample discussed below.

\begin{figure}
   \centering
 \includegraphics[width=78mm, height=75mm]{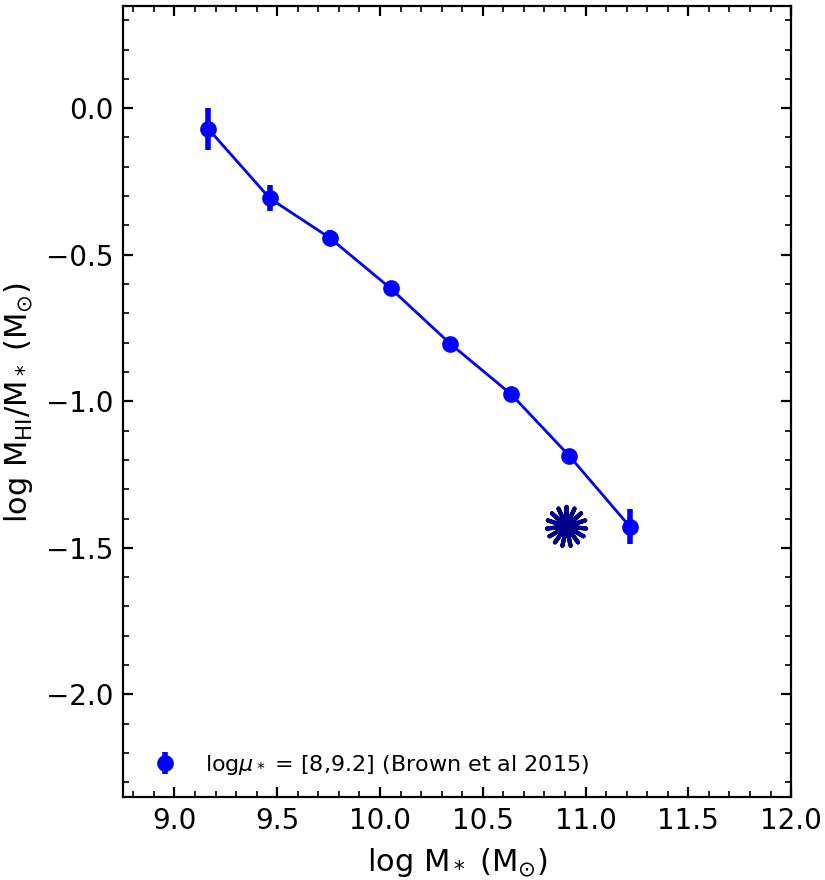}
    \caption{The average stacked H\textsc{i} fraction as function of the stellar mass \citep{Brown2015}. In blue we show the relation separated into galaxies with stellar surface brightness comparable to that of JO206 within a factor of 4, the error bars represent the scatter in the mass bins. JO206 is illustrated by the blue asterisk.}\label{HIstack}
 \end{figure}

Figure\,\ref{HIstack} shows the relation of the \HI\ fraction as a function of stellar mass for galaxies in which $\mu_{\ast}$ is the same as that of JO206 within a factor of 4. This factor is much larger than the aforementioned systematic error on $\mu_{\ast}$ caused by using the $K$-band rather than $z$-band R$_{50}$ radius. JO206 is indicated by the blue asterisk on this relation. A comparison of the measured stellar surface density of JO206 with its counterparts between log$\mu_{\ast}$ = 8.0  to 9.2, places it about 0.3\,dex below the average \HI\ fraction. Based on this scaling relation the galaxy appears to be \HI\ deficient and missing about 50 percent of its expected \HI\ gas mass compared to galaxies with the same M$_{\ast}$ and $\mu_{\ast}$. This is consistent with the gas loss estimated as described section\,\ref{estGasFrac}.
\\\

\subsection{HI Displacement}\label{HImodel} 
Having established that the galaxy has lost about 50 percent of its initial \HI, Fig.\,\ref{HIoverVband} shows that most of the remaining \HI\ is not distributed on a settled disc. In this section we aim to quantify the displacement relative to the galaxy disc in order to establish how much of the \HI\ is at larger radii than expected for an \HI\ disc with the same mass (i.e., how much \HI\ is in the tail). To do so we model the \HI\ distribution assuming no ram-pressure stripping is affecting the galaxy and that all of its observed \HI\ content is intact. 

The model is based on the well-established and tight correlation that exists between \HI\ masses and diameters of galaxies, and the self-similarity of the \HI\ radial profiles (\citealp{Broeils1997}, \citealp{Verheijen2001}, \citealp{Noordermeer2005}, \citealp{Begum2008}, \citealp{Wang2016}, \citealp{Martinsson2016}).  

We use the \HI\ size-mass relation parametrised by \citet{Wang2016} as, 
\begin{equation}
\mathrm{log(D_{HI}/kpc) = 0.51 log(M_{HI}/M_{\odot}) -  3.32}, 
\end{equation}

where the \HI\ diameter, D$_{\rm HI}$ is defined at the \HI\ surface density of 1 \MSUN\ pc$^{-2}$ (N$_{\rm HI} = 1.25 \times 10^{20}$ atoms\,cm$^{-2}$). The observed scatter about this relation is 0.06 dex. 

Based on this relation we calculate an \HI\ diameter of D$_{\rm HI}$  = 33\,kpc for the unperturbed JO206 model. The radial \HI\ surface density distribution was determined using the \citet{Martinsson2016} \HI\ profile formulated as, 

\begin{equation} 
\mathrm{\Sigma_{\rm H\textsc{i}}(R) = \Sigma_{\rm H\textsc{i}}^{max}.e^{\frac{-(R - R_{\Sigma,max})^{2}}{2\sigma_{\Sigma}^{2}}}}
\end{equation} 
In this equation the parameters R$_{\Sigma,max}$ and $\sigma_{\Sigma}$ are fixed to the values 0.2D$_{\rm HI}$ and 0.18D$_{\rm HI}$, respectively \citep{Martinsson2016}. The only free parameter is $\Sigma_{\rm H\textsc{i}}^{max}$, which we set to 0.4 \MSUN pc$^{-2}$ such that $\Sigma_{\rm HI}$(D$_{\rm HI}$/2) = 1 \MSUN pc$^{-2}$.

\begin{figure}
 \includegraphics[width=84mm, height=43mm]{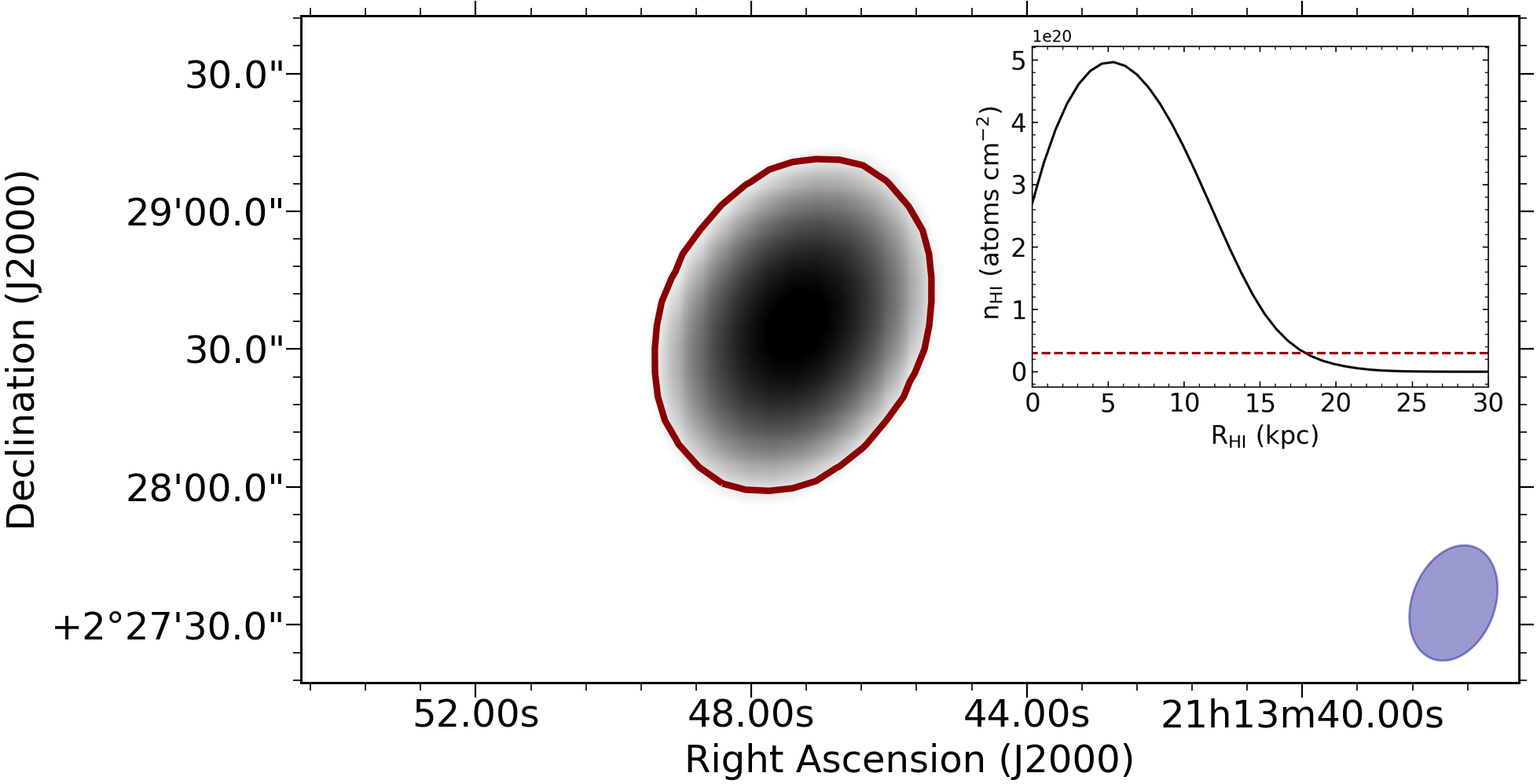}\\
  \includegraphics[width=84mm, height=44mm]{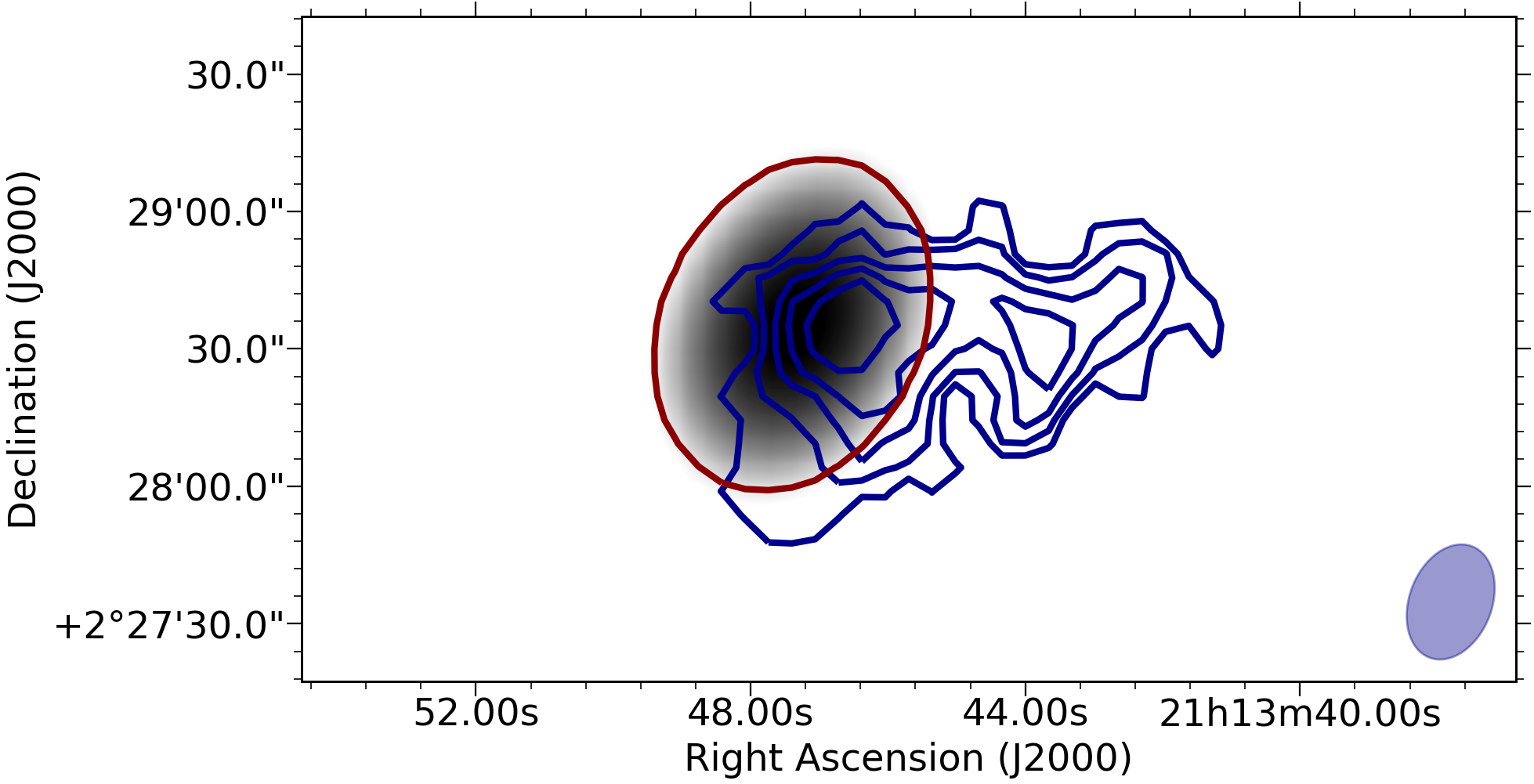}
    \caption{The top panel shows the simulated H\textsc{i} distribution. The red contour is drawn at a column density of  $3 \times10^{19}$ atoms cm$^{-2}$, the same as the lowest H\textsc{i} contour in our data. The red ellipse has a semi-major axis of 28\,kpc. In the inset panel we show the model input H\textsc{i} radial profile with the red horizontal line indicating our sensitivity limit. In the bottom panel we compare the model H\textsc{i} with the observed distribution (blue contours). The observed H\textsc{i} column density levels are $n_{\rm H\textsc{i}}$ = 3, 6, 9, ... $\times$ 10$^{19}$ atoms/cm$^{2}$. The FWHM beam size of $26\arcsec \times 18\arcsec$ is indicated by the blue ellipse.}\label{modelHI}
 \end{figure}

With the above-mentioned tools, and assuming the observational conditions described in \cref{sec:vlaObs}, we modelled the unperturbed \HI\ distribution of JO206 using the 3D-Barolo package \citep{DiTeodoro2015}. We used the position angle and inclination of the stellar disc as the inputs for model. The model was convolved with the \HI\ beam within 3D-Barolo. The resulting moment-0 map with the radial \HI\ surface density profile is shown in the top panel of Fig.\ref{modelHI}. The slight misalignment between the \HI-model and stellar disc is attributed to the effect of beam smearing by the \HI-beam which dominates this model.

The red ellipse in the figure outlines the model \HI\ disc with a radius, R$_{\rm HI}$=28\,kpc as defined at the \HI\ column density sensitivity of $3 \times 10^{19}$ atoms cm$^{-2}$. This corresponds to the sensitivity of our VLA observations and to the lowest \HI\ contour shown for JO206 in Fig.\,\ref{HIoverVband}. The model \HI\ outside the red contour would not have been detected by our VLA observations and is therefore not considered here.
We compared the model and observed moment-0 maps as shown in bottom panel of Fig.\,\ref{modelHI}. Any \HI\ emission in the observed map that lies outside of the red contour should not be there and is considered the \HI\ tail. Within this tail the measured \HI\ mass is $1.8 \times 10^{9}$ \MSUN, which is $\sim$ 60\% of the total \HI\ mass in JO206. 

We also determined how much of the  total \HI\ mass has simply been displaced relative to the model, regardless of whether it is inside or outside the $3 \times 10^{19}$ atoms cm$^{-2}$ model isophote in Fig.\,\ref{modelHI}. This is calculated as; 
\HI$_{\rm disp}$ $=$ $\Sigma|$$mod$$_{i,j} -$ $obs$$_{i,j}|/2$, where $mod$ is the modelled \HI, $obs$ is the observed, and the indices $i,j$ run through all pixels in the image. Using this formulation we find that a \HI\ mass of $2.0 \times 10^{9}$ \MSUN\ has been displaced relative to the model with about half of the signal found in the tail and the other half in the disc.

\section{Gas and Star Formation Activity}\label{coldSFR}
\begin{figure}
   \centering
 \includegraphics[width=75mm, height=72mm]{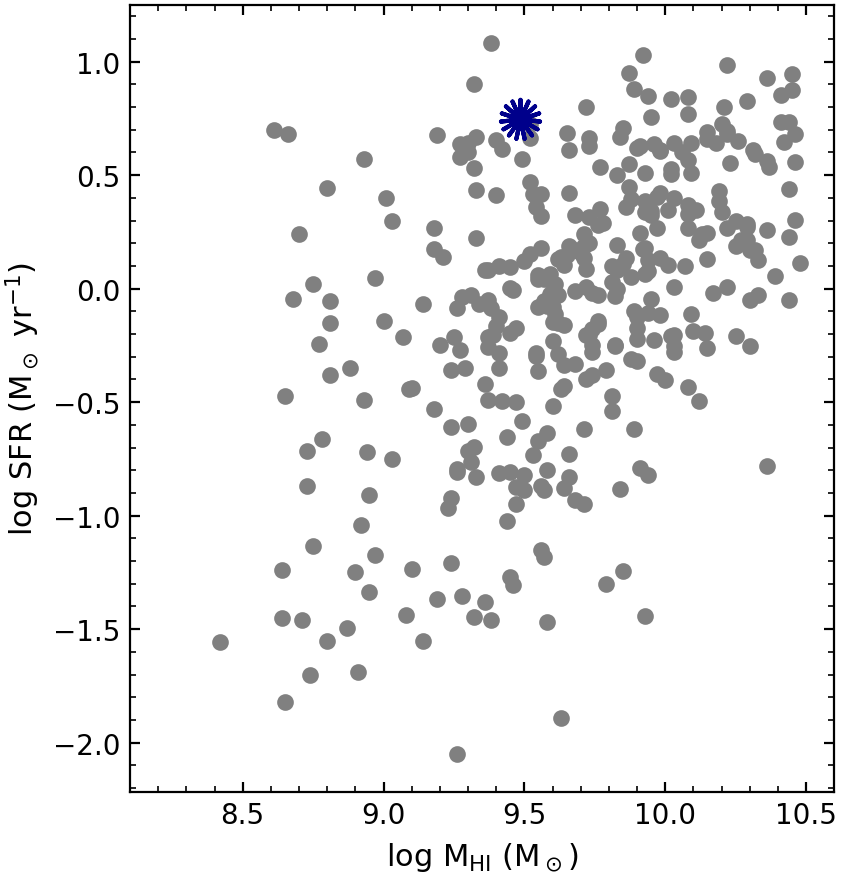}
    \caption{The star formation rate as function of H\textsc{i} mass for galaxies with the same stellar masses as JO206. Grey points are galaxies from the GASS and the blue asterisk represents the JO206 galaxy.}\label{msSFR}
 \end{figure}

We have determined that JO206 has lost about half of its neutral gas due to ram-pressure by the ICM, and that 60 per cent of its remaining \HI\ content is distributed along the 90\,kpc-long tail of stripped gas. In this section we study the galaxy's star formation efficiency under these stripping conditions by examining the relation between \HI\ content and star formation rate. Given the known correlation between star formation rate and stellar mass (e.g., \citealp{Brinchmann2004}, \citealp{Salim2007}) we start by selecting a sample of galaxies with the same stellar mass as JO206. To do so we extracted all galaxies in the GALEX Arecibo SDSS Survey (GASS; \citealp{Catinella2010}) with a stellar mass within a factor of 4 of that of JO206. Then since star formation rate scales with \HI\ mass at a fixed stellar mass (\citealp{Doyle2006}, \citealp{Saintonge2016}, \citealp{Huang2012}), we plot star formation rate vs \HI\ mass for this control sample in Fig.\,\ref{msSFR}. We find that in comparison JO206 (SFR = 5.6 \MSUN yr$^{-1}$) is about 0.5\,dex above the general galaxy population implying that it is forming more stars than similar galaxies given its \HI\ mass. The \HI-SFR scaling relation has a large scatter, as shown in the aforementioned papers and in Fig.\,\ref{msSFR}. Therefore, the exact SFR enhancement has large uncertainties. However, the fact that JO206 lies at the very edge of the observed distribution of our comparison sample is a relatively strong indication that some SFR enhancement has occurred. We calculate an overall \HI\ gas depletion time scale as $\tau_{d}$  = M$_{\rm HI}$/SFR, and find that it is 0.54 Gyr. This time scale is shorter than that of a typical normal disc galaxy ($\sim$2 Gyr; \citealp{Leroy2008}). This indicates that the star formation rate is enhanced in this object compared to other galaxies with similar stellar and \HI\ masses. The total SFR measured for JO206 would be more typical for a galaxy with the same stellar mass and an \HI\ mass of approximately $3 \times 10^{10}$ \MSUN, an order of magnitude above the measured M$_{\rm HI}$ of this galaxy. 

\subsection{Comparing H$\alpha$ to HI } 
\begin{figure*}
 \includegraphics[width=120mm, height=130mm]{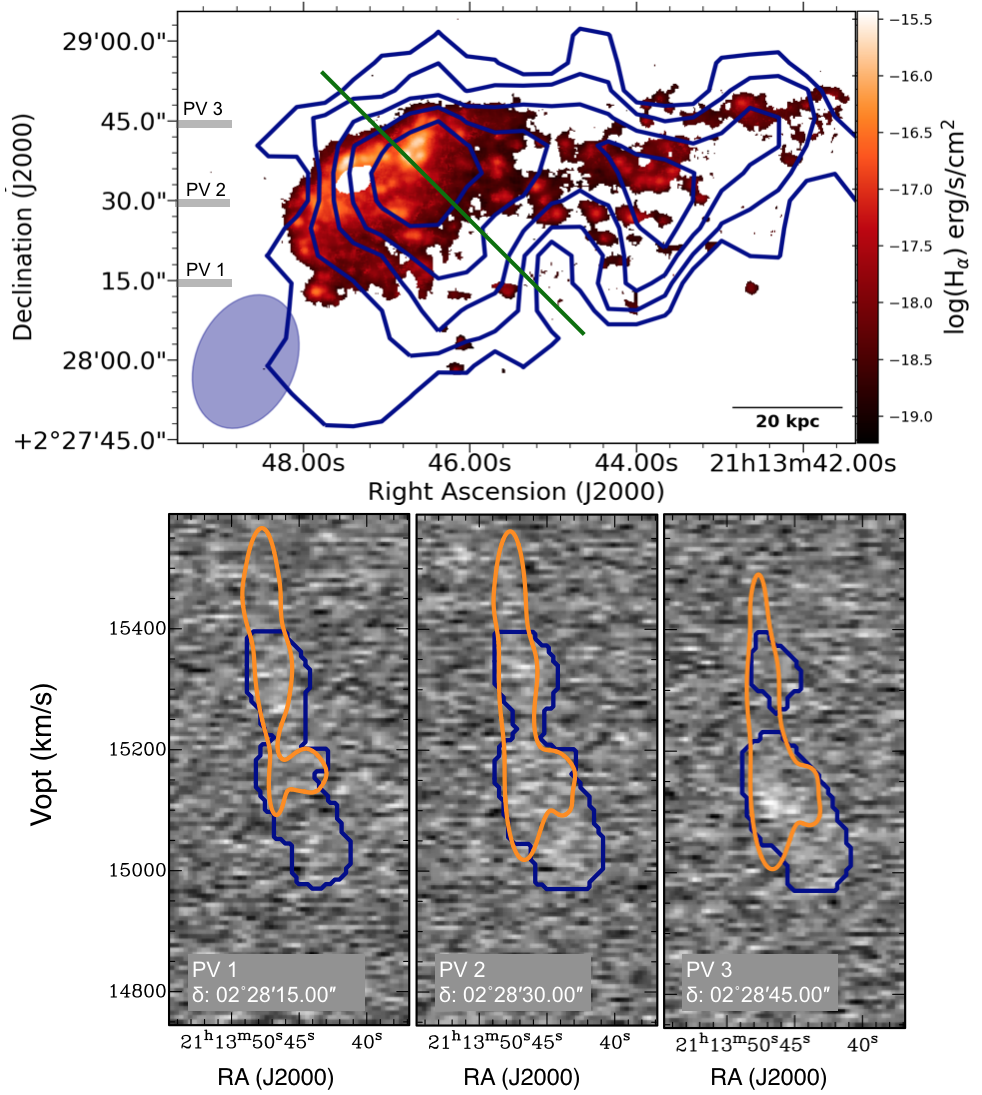}
 \caption{A comparison of the H\textsc{i} column density distribution with H$\alpha$ map from MUSE. In the top panel the MUSE H$\alpha$ is shown in red with the H\textsc{i} distribution overlaid in blue at column densities of 3, 6, 9, ... $\times$ 10$^{19}$ atoms/cm$^{2}$. The green diagonal line indicates the slit position for the profile in Fig.\,\ref{ProfileHIandHalpha}. The bottom panel shows position velocity diagrams (PVDs) extracted along three declination slices (PV1, PV2 and PV3) in the H\textsc{i} cube. Velocities in the PVDs are in the optical definition using the barycentric standard-of-rest. The H$\alpha$ emission convolved with the H\textsc{i} PSF is illustrated in orange and the underlying H\textsc{i} mask in the image cube is outlined by the blue contours.}\label{PVDs}
 \end{figure*}

\begin{figure}
  \includegraphics[width=85mm, height=48mm]{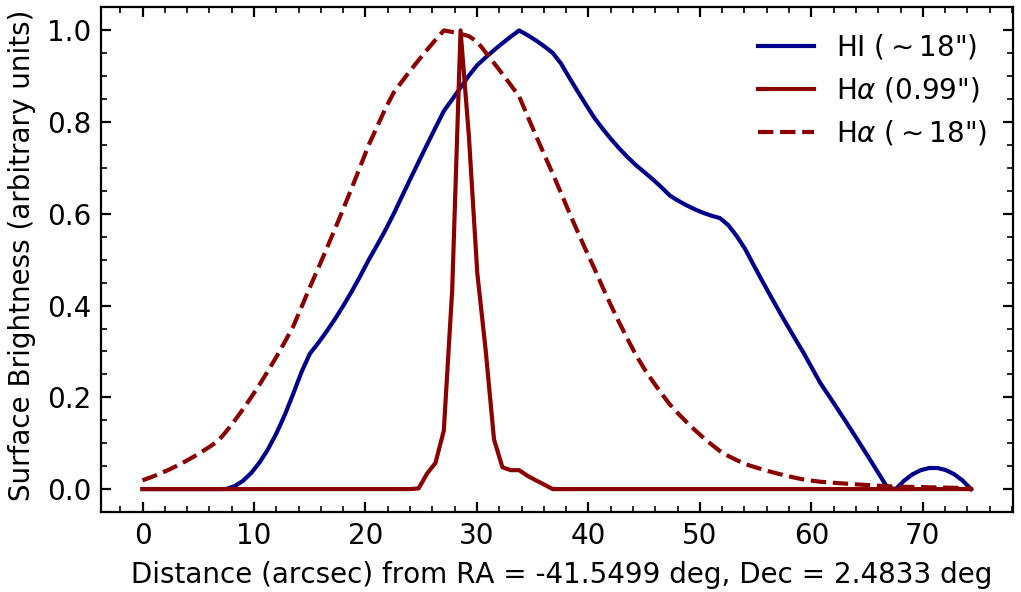}
    \caption{Comparison profiles along the diagonal line in Fig.\,\ref{PVDs}. The red solid line represents the full resolution ($\sim$0.99\arcsec) H$\alpha$ profile and the red dashed line is the H$\alpha$ profile convolved with the H\textsc{i} PSF ($\sim$18\arcsec). Shown in blue is the H\textsc{i} line emission. The y-axis has arbitrary units and the x-axis is the distance in arcsec from the starting point (RA = -41.5499 deg, Dec = 2.4833 deg) of the line from which the profile was extracted (i.e., the green diagonal line in Fig.\ref{PVDs}).}\label{ProfileHIandHalpha}
 \end{figure}

To understand the reason for the higher than expected star formation activity of JO206 we use H$\alpha$ map (dust- and absorption-corrected) as a tracer for recent star formation. We compare its distribution with that of \HI\ as shown in Fig.\,\ref{PVDs}. The \HI\ emission overlaps almost entirely with H$\alpha$. The disagreement seen is due to the different spatial resolutions between the \HI\ and H$\alpha$ data. We demonstrate this in Fig.\,\ref{ProfileHIandHalpha} by comparing \HI\ and H$\alpha$ surface brightness distribution along an arbitrary axis line (solid green line in Fig.\,\ref{PVDs}) both before and after convolving the H$\alpha$ image with the \HI\ PSF. Fig.\,\ref{ProfileHIandHalpha} shows that H$\alpha$ emission is found at all positions with \HI\ emission after convolution.

The \HI\ map exhibits two local maxima where it peaks at \HI\ column densities of $\sim1.6 \times 10^{20}$ atom/cm$^{2}$. One is within the galaxy disc while the other is located at $\sim$60\,kpc from the galaxy centre in the middle of the tail. These bright \HI\ regions coincide with areas with several knots of H$\alpha$ emission. These are regions at which we expect recent star formation. However, in the southern pointing (see Appendix A, Fig.\,\ref{newMUSEpointing}) the \HI\ is much more extended than the H$\alpha$ knots but it is at such low column densities that no star formation is expected in those regions.

The comparison between the H$\alpha$ and \HI\ position velocity diagrams in the bottom panel of Fig.\,\ref{PVDs} shows a general close agreement in the kinematics as well. However, the H$\alpha$ appears to have a slightly higher velocity than the \HI. This is due to the presence of H$\alpha$ but not of \HI\ (at the same angular resolution) in the eastern edge of the galaxy disc, which has receding velocity compared to systemic (see figures 6 and 7 in Poggianti et al. 2017b). The agreement between the velocities of cold \HI\ gas and H$\alpha$, particularly along the tail is attributed to the ionisation of the stripped gas by newly forming stars.

 \subsection{Star Formation Density and HI}
\begin{figure}
 \includegraphics[width=88mm, height=48mm]{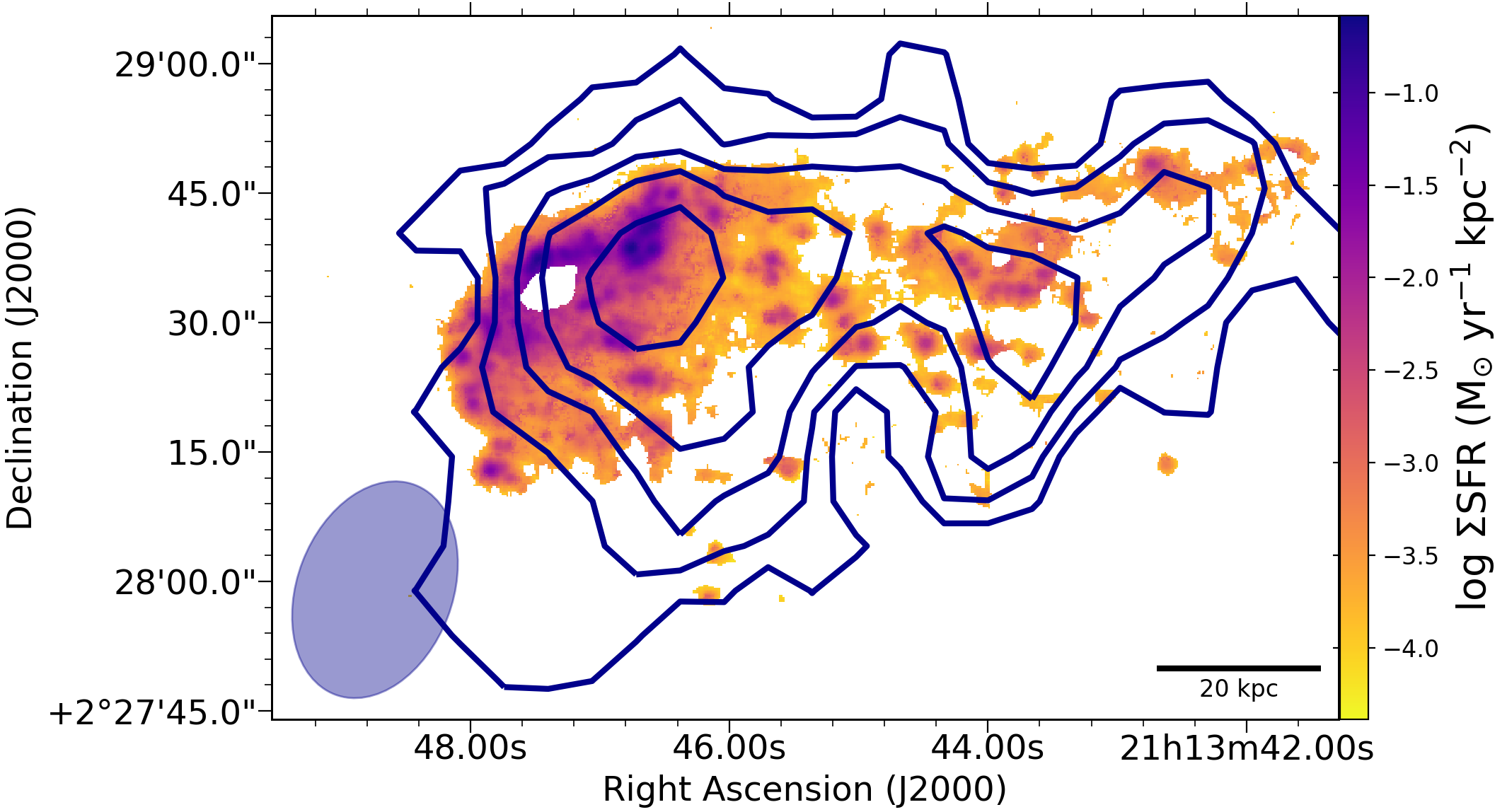}
    \caption{The star formation rate surface density map with the H\textsc{i} contour levels at the same levels as in Fig.\,\ref{HIoverVband}. The white patch is the location of the excluded central AGN.}\label{SFRDmap}
 \end{figure}

To pinpoint where the observed enhanced star formation activity is taking place in JO206 we examine its star formation rate in relation to the \HI\ gas density distribution. We start globally by comparing the star formation rate density map and the \HI\ distribution in Fig.\,\ref{SFRDmap}. 

The map excludes H$\alpha$ emission from the AGN. It shows an increased star formation rate within the disc in the western region around 40\,kpc from the centre of the disc. This coincides with the location of the bright \HI\ emission where CO has also been detected (see Fig.\,\ref{COMap}; also \citealp{Moretti2018}). Further west in the tail there is continued star formation activity spread out in various regions with a star forming knot around 60\,kpc from the centre again corresponding with where the \HI\ emission peaks. The APEX observation here can only provide an upper limit of $6.3 \times 10^8$ \MSUN\ of molecular gas \citep{Moretti2018}.  

The comparison shown in Fig.\,\ref{SFRDmap} however only offers a qualitative assessment of the star formation rate and gas correlations. It gave us clues to where the galaxy is likely forming stars with high efficiency. 

We take this investigation further by studying this correlation in a spatially resolved way. This approach allows for improved statistics to pinpoint exactly where in the galaxy (disc or tail), gas forms stars efficiently. The effectiveness of this approach was illustrated by \citet{Bigiel2008} and \citet{Bigiel2010} who found a higher star formation efficiency in the inner regions of ``normal'' field spiral galaxies compared to the outer parts, the latter being more likely dominated by \HI. \citet{Boissier2012} also applied this approach to study star formation efficiency of a sample of galaxies that are possibly experiencing ram-pressure gas stripping in the Virgo cluster. They found that star formation rate was lower by an order of magnitude in the gas stripped tails than within the galaxy. 

To make a direct comparison between the star formation and \HI\ map, we convolved the star formation surface density map (Fig.\,\ref{SFRDmap}) with the \HI\ beam and regridded to the same pixel scale of 5\arcsec. During this process the flux in the star formation surface density map was conserved. We then flagged pixels as being in the tail or disc based on whether they are inside or outside the \HI\ model's 3 $\times$ 10$^{19}$ atoms/cm$^{2}$ contour (see \cref{HImodel}) and performed a pixel-by-pixel analysis similar to the aforementioned studies.

As a control sample we use ``normal'' field galaxies based on The HI Nearby Galaxy Survey (THINGS; \citealp{Bigiel2008}, \citealp{Bigiel2010}). Although star formation rate surface density maps of the control sample are based on a combination of FUV and 24$\mu$m emission, a good agreement was shown with the H$\alpha$ emission for this sample (see Fig\,4 in  \citealp{Bigiel2008}). This makes this sample suitable to compare with our measurements. Image maps in this control sample have a 750\,pc spatial resolution. To enable a direct comparison with JO206 measurements we convolved these images to the same physical resolution as JO206 using a circular PSF with the same area as our $26\arcsec \times 18\arcsec$ \HI\ beam which is $\sim$22\,kpc at the distance of JO206. We then only selected a control subsample of five galaxies which remained sufficiently resolved after this convolution, namely NGC\,5055, NGC\,2841, NGC\,7331, NGC\,3198 and NGC\,3521. 	

\begin{figure}
   \centering
 \includegraphics[width=83mm, height=70mm]{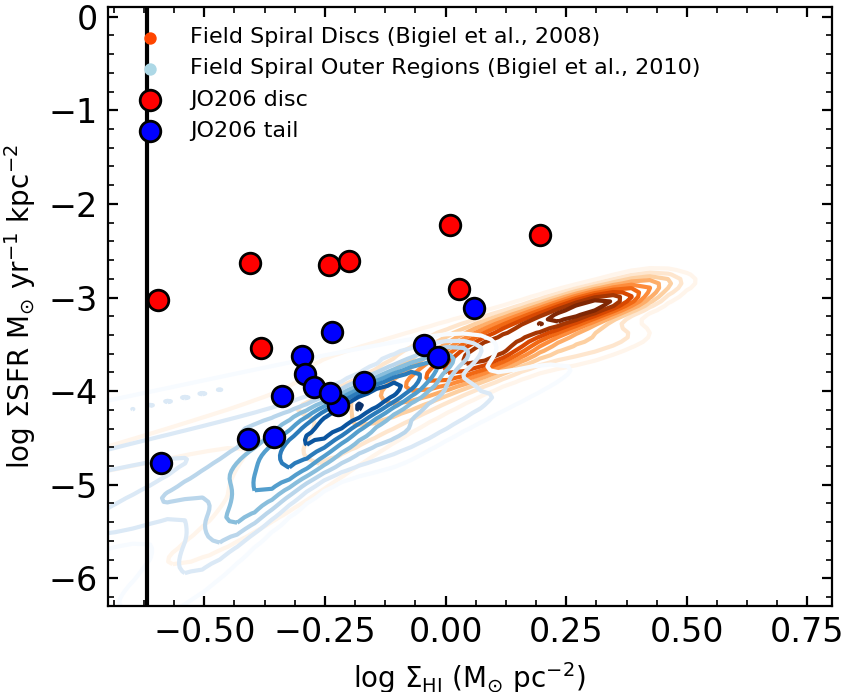}
    \caption{Relation between the star formation rate density and H\textsc{i} surface density. The red and blue points represent the main galaxy body and tail of JO206, respectively. These are plotted independently per beam. Orange density contours are the inner regions (discs) of field spiral galaxies selected from the THINGS sample from \citealp{Bigiel2008} convolved with the H\textsc{i} beam. Light blue density contours represent the outer regions of spiral galaxies in the field \citep{Bigiel2010} also convolved with the the H\textsc{i} beam. Contrary to the high resolution map, THINGS galaxies form a single sequence. The solid vertical line indicates our H\textsc{i} sensitivity limit. }\label{pixelSFR}
 \end{figure}

Figure\,\ref{pixelSFR} shows the resolved star formation rate surface density versus H\textsc{i} surface density within the disc (red) and in the tail (blue) of JO206. This is compared with the inner and outer regions of the THINGS control subsample smoothed to our resolution as described above. We find that the inner parts of JO206 are producing stars at a relatively higher rate than the outer regions at a given \HI\ surface density.  The star formation rate is about 10 times higher on average in the inner parts of the galaxy than the outer regions. This is not surprising since the stripped \HI\ is less dense in the tail compared to the disc. However, both the inner and outer pixels of JO206 show higher star formation rates for a given \HI\ surface density compared to normal field galaxies in the control sample. 

\begin{figure}
   \centering
 \includegraphics[width=85mm, height=63mm]{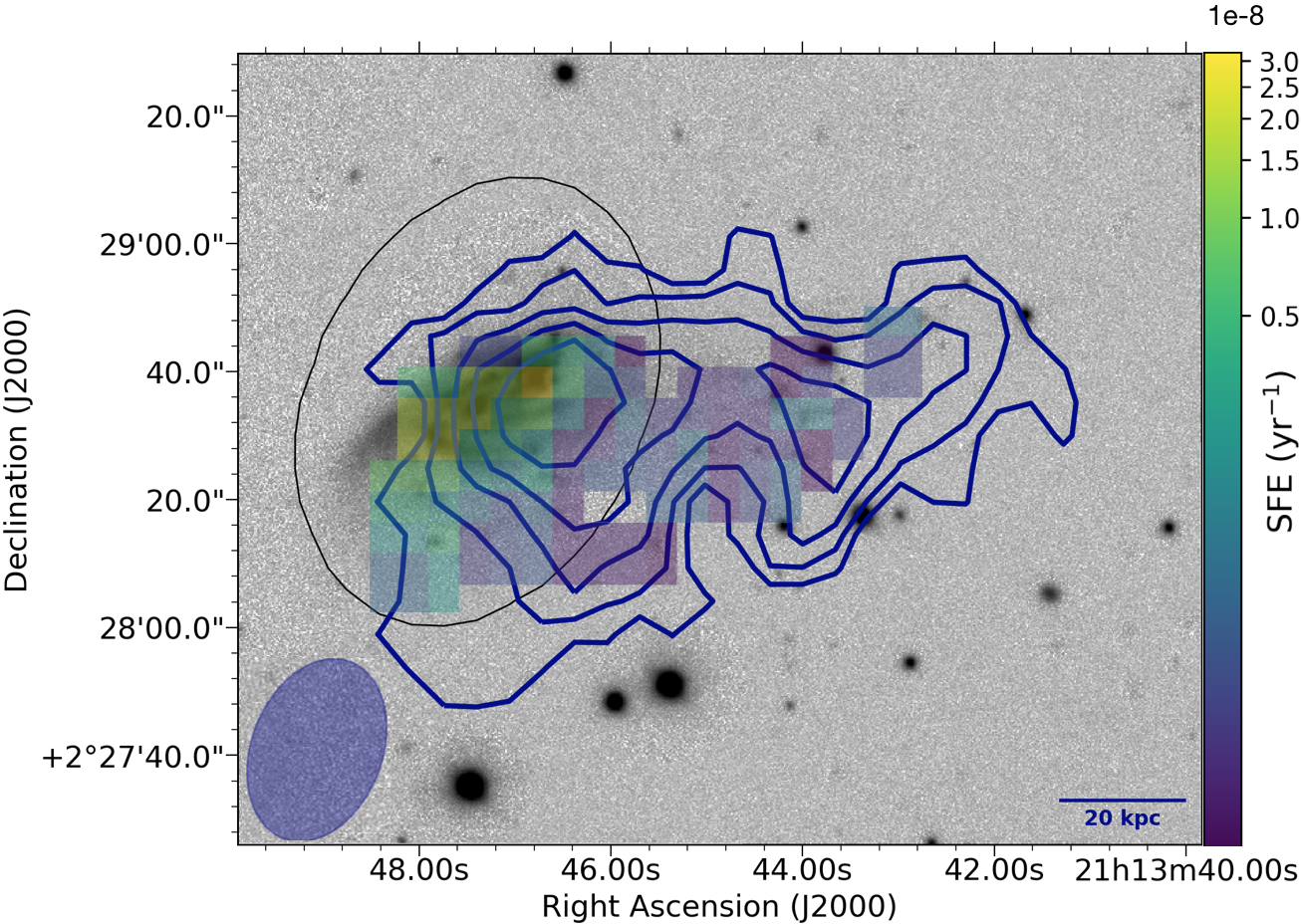}
    \caption{The star formation efficiency (SFE) map with the colour bar indicating SFE values for pixels with SFE $\geq 10^{-9}$ yr$^{-1}$. The map is  overlaid over the optical V-band image of JO206 and its H\textsc{i} distribution. H\textsc{i} column densities are $n_{\rm H\textsc{i}}$ = 3, 6, 9, ... $\times$ 10$^{19}$ atoms/cm$^{2}$. Outlined in black is the H\textsc{i} disc defined at the H\textsc{i} model's 3 $\times$ 10$^{19}$ atoms/cm$^{2}$ contour.}\label{SFEpix}
 \end{figure}

In total we measure an average total Star Formation Efficiency (SFE; measured as SFR/M$_{\rm HI}$) of  $\sim$2 $\times$ 10$^{-9}$ yr$^{-1}$ which is about a factor of 5 times higher than normal field local galaxies (\citealp{Schiminovich2010}, \citealp{Leroy2008}). We show in Fig.\,\ref{SFEpix} the locations at which the JO206 is most efficient (SFE $\geq 10^{-9}$ yr$^{-1}$) at forming stars. Note that some correlation between adjacent pixels in the SFE map is present.

The observed enhanced star formation activity in the disc and especially the tail of JO206 persists even when compared to other ram-pressured gas stripped galaxies in the Virgo cluster (see Fig.\,3 by \citealp{Boissier2012}). This could be because JO206 is a much more extreme case of ram-pressure stripping based on the extreme length of the observed tail. 
Furthermore, this spatially resolved result confirms the more global results of enhanced star formation activity in JO206 based on just \HI\ mass and star formation rates values (see Fig.\,\ref{msSFR}). 

\section{Summary}\label{summary} 
As part of the ESO MUSE GASP survey we have studied the \HI\ gas phase of the prototypical ``jellyfish'' galaxy in the sample, namely, JO206 \citep{Poggianti2017}. This massive galaxy (M$_{\ast} \sim$ $8.5 \times10^{10}$ \MSUN) resides in a poor cluster but exhibits a long H$\alpha$ tail caused by ram-pressure stripping. In this paper we primarily focused on the stripping of \HI\ and on the star formation activity associated with its \HI-content using data obtained with the VLA telescope. Our finding are as follows;

\begin{itemize}
\item{As the result of ram-pressure stripping, we find that the \HI\ distribution is perturbed and exhibits a one-sided, $\sim$90\,kpc \HI\ tail from the optical disc. We measure a total \HI\ mass of M$_{\rm HI} = 3.2 \times 10^{9}$ \MSUN\ of which 60\% ($1.8 \times 10^{9}$ \MSUN) is in the gas stripped tail. Overall the galaxy is about 50 per cent \HI\ deficient. The observed stripped fraction of \HI\ gas is consistent with estimations of gas mass lost from modelling of the cluster's ram-pressure and the galaxy's restoring force.} \\

\item{ An assessment of the \HI\ and star formation rates shows that the galaxy is generally undergoing an enhanced star formation activity compared to its counterpart with the same stellar mass. The \HI\ depletion time is $\sim$0.5 Gyr in this galaxy which is shorter than that of ``normal'' spiral galaxies in the field. }\\

\item{By comparing the \HI\ distribution and the H$\alpha$ distribution, we find a strong correlation between the observed cold gas and ionised emission (H$\alpha$). This is seen in both the galaxy main body and the tail. Moreover, the agreement persists in the kinematics of the galaxy. This indicates a strong link between the presence of cold gas and the recent star formation across all of the galaxy.} \\

\item{To pinpoint the exact locations at which the new stars are forming with high efficiency, we smoothed and regridded the MUSE SFRD maps to our \HI\ resolution and pixel scale, and conducted a pixel-by-pixel analysis of the star formation rate density and the \HI\ surface density. We tagged pixels belonging to the galaxy disc and tail.  Our results show that the star formation efficiency in the disc is on average $\sim$10 times higher compared to the tail for a given \HI\ surface density. We find that in general the inner and outer parts of JO206 have relatively higher star formation efficiencies compared to galaxies in the literature (\citealp{Bigiel2008}, \citealp{Bigiel2010}), even compared to those undergoing ram-pressure stripping in the Virgo cluster \citep{Boissier2012}.}

\end{itemize}

This work contributes to the ongoing efforts by the GASP survey in understanding the fate of gas in galaxies during the ram-pressure stripping phenomenon. We highlighted the importance of studying the  \HI\ gas phase and understanding the link between the gas and star formation activity of the JO206 galaxy as it falls into the cluster. The GASP data showed that star formation in the gas stripped tails is common in the sample.  However, understanding particularly the formation of stars in the gas stripped tails is not trivial and this is the case with JO206. There are a number of properties to be taken into account such as the environmental conditions. JO206 is special in this case because it is massive galaxy undergoing ram-pressure in a rather poor galaxy cluster. Although a significant fraction ($\sim$60\%) of the \HI\ has been displaced, it is still visible in the form of \HI. i.e., it has not been immediately ionised by (or lost to) the ICM. In other words, at its current stripping stage, JO206 still has fuel to form new stars along its tail and disc. Comparing this galaxy with others in the GASP sample in different environments will clarify whether the environment played a pivotal role in the enhanced observed star formation or whether other specific physical conditions are responsible.

\appendix
\section{Supplementary MUSE observations}
VLA observations detected \HI\ emission outside the region covered by the initial GASP observations with MUSE presented in GASP I \citep{Poggianti2017}, as shown in Fig.\,\ref{HIoverVband}. To check for possible H$\alpha$ emission in the southern tail of the \HI\ emission we obtained new MUSE observations. These were carried out as part of ESO programme 0102.C-0589, a filler program designed to use idle time at the VLT UT4 during the worst weather conditions. Observations were carried out under non-photometric conditions between October 3rd and October 6th 2018; we obtained $16\times900$ seconds exposures of a single MUSE pointing. The total exposure time of 4h is much larger than the one of normal GASP observations (2700s) to compensate for possible loss due to the bad weather conditions. 

Observations were carried out under thin and thick cirrus and seeing between 1.3 and 0.6 arcsec (as measured by the DIMM).
The new MUSE data were reduced using the standard GASP data reduction procedure (see GASP I \citet{Poggianti2017} for details).
Because the observations were carried out under non-optimal weather conditions, we corrected the fluxes by comparing the H$\alpha$ fluxes measured in the regions in common with the original GASP observations. We found that the fluxes obtained from the new observations are 1.66 times lower than those obtained from the original GASP data. We therefore re-scaled the data cube obtained from the new observations by this factor to put all observations on the same flux scale.

\begin{figure}
   \centering
 \includegraphics[width=85mm, height=63mm]{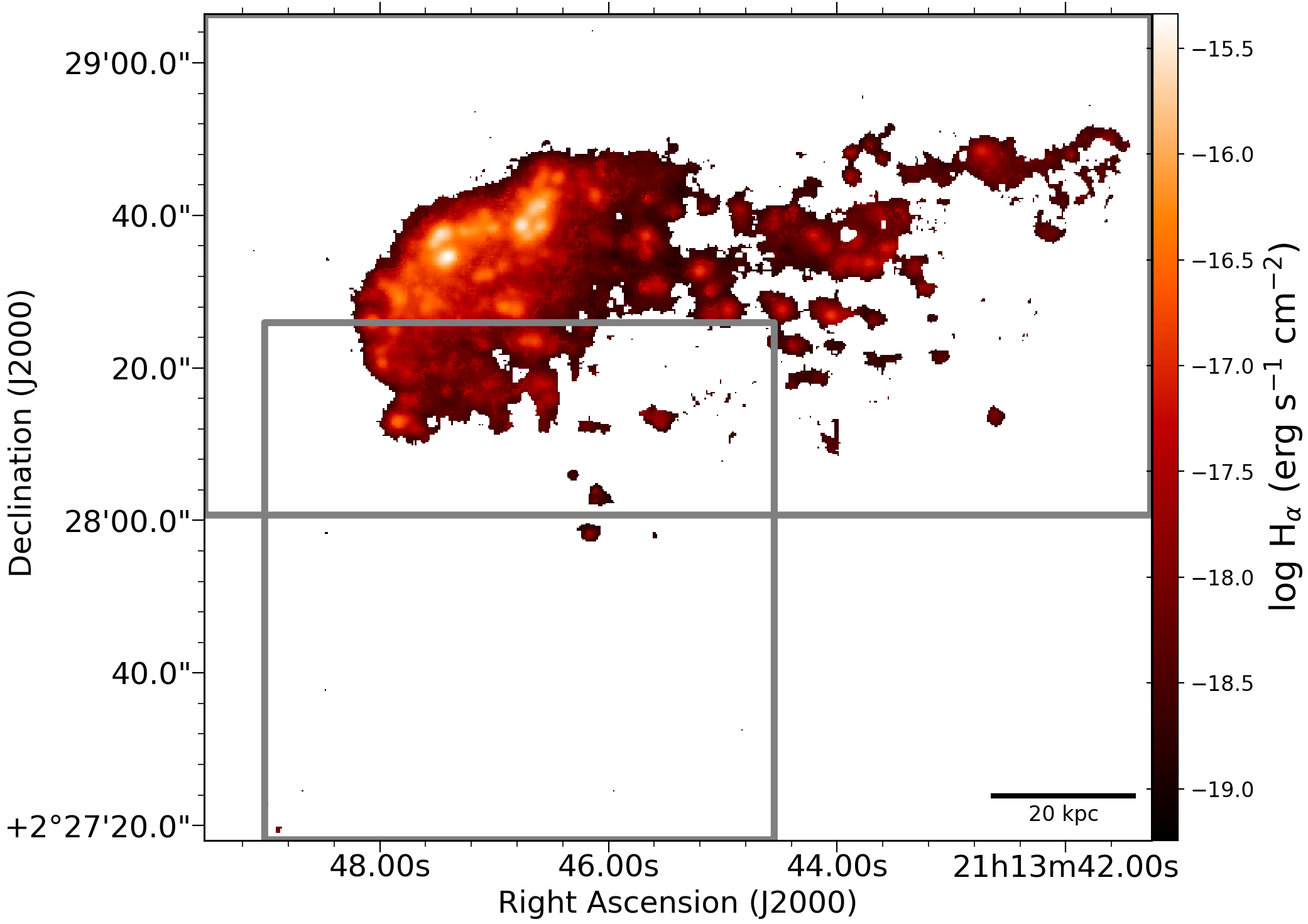}
    \caption{The MUSE H$\alpha$ map with the old and new MUSE FOV pointings overlaid are outlined in grey. The new pointing to the south of the map.}\label{newMUSEpointing}
 \end{figure}

The combined H$\alpha$ emission map is shown in Fig.\,\ref{newMUSEpointing} together with the FoV of the new MUSE observations and of the original GASP ones. We compared the r.m.s. of the background and the distribution of the measured H$\alpha$ fluxes in the overlapping regions and we conclude that the new observations are at least as deep as the original GASP ones. The new observations revealed three new H$\alpha$ blobs beyond the region covered by the original GASP observations. An analysis of the BPT diagram indicate that the gas in these three regions is ionised by young stars.

\newpage
\section*{Acknowledgements}
We would like to thank the anonymous referee for the insightful comments that significantly improved the quality of the analysis and presentation of the paper. We thank Toby Brown for his invaluable help with the scaling relations. This project has received funding from the European Research Council (ERC) under the European Union's Horizon 2020 research and innovation programme (grant agreement no. 679627, project name FORNAX). We acknowledge funding from the INAF PRIN-SKA 2017 program 1.05.01.88.04 (PI L. Hunt). YJ acknowledges financial support from CONICYT PAI (Concurso Nacional de Insercion en la Academia 2017), No. 79170132 and FONDECYT Iniciaci\'{o}n 2018 No. 11180558. MV acknowledges support by the Netherlands Foundation for Scientific Research (NWO) through VICI grant 016.130.338. This work is based on observations collected at the European Organisation for Astronomical Research in the southern hemisphere under ESO programme 196.B-0578 (PI B.M. Poggianti) and  programme 0102.C-0589 (PI F. Vogt). This work made use of THINGS, 'The \HI\ Nearby Galaxy Survey' \citep{Walter2008}. This paper makes use of the VLA data (Project code: VLA/17A-293). The National Radio Astronomy Observatory is a facility of the National Science Foundation operated under cooperative agreement by Associated Universities, Inc.

\normalsize
\bibliographystyle{mn2e.bst} 
\bibliography{J206GaspRef.bib} 

\end{document}